\documentstyle{article}
\title{Coherent states of quantum non-linear systems}
\author{Z. Haba\\Institute of Theoretical Physics,University of
Wroclaw,Wroclaw,Poland}
\date{}
\begin{document}
\maketitle
\begin{abstract}
Quantum dynamics of integrable systems is discussed.
 Localized wave packets
  generalizing the conventional
coherent states of minimal uncertainty are constructed. The wave packet
moves along a certain trajectory and does not change its shape
for times of order $\frac{1}{\hbar}$.
                                   \end{abstract}
In this Letter we suggest that the definition of a coherent state should be
related to the Hamiltonian.The conventional coherent states \cite{Glau}
are defined as eigenstates of the annihilation operator
\begin{equation}\alpha\mid z> =z\mid z> \end{equation}
The real and the imaginary parts of the complex variable
z can be expressed by the mean values of the position $\hat{x}$ and
momentum  $\hat{p}$ operators in the coherent state. Under the oscillator
Hamiltonian evolution $U_{t}$
  \begin{equation}
 U_{t}|z>=|exp(-i\omega t) z>
\end{equation}
where $\omega$ is the oscillator$^{\prime}$s frequency.
Then, because the state $\mid z> $ is concentrated around
$x=\sqrt{\frac{2\hbar}{\omega}}Rez$
 it follows
that $U_{t}|z>$ has its support on the classical trajectory
$x(t)=\sqrt{\frac{2\hbar}{\omega}}Re(exp(-i\omega t)z)$.
The simple classical evolution $z\rightarrow exp(-i\omega
t)z$ is a consequence of the simple representation of
$\sqrt{\omega}x+i\frac{p}{\sqrt{\omega}}=\sqrt{I}exp(i\theta)$ in terms
 of the action-angle variables $(I,\theta)$ for the harmonic oscillator.

We are going to generalize the construction of coherent
states to integrable non-linear systems.
It is known that the quantum non-linear dynamics cannot be
transformed into a coherent state dynamics exactly (see the discussion in
\cite{niet1}\cite{niet5};these authors define generalized coherent states,
which
do not coincide with ours).
However, wave packets approximately localized around the classical
trajectory of an electron in the hydrogen atom have been constructed
\cite{most}\cite{naum} .
For this purpose a large dynamical symmetry relating the hydrogen
atom to the fourdimensional oscillator has been utilized.
In this Letter we do not assume any symmetry, but the integrability of
a classical Hamiltonian system. Our construction  of wave packets
can be applied to multidimensional  anharmonic
oscillators. The localized states can describe a classical behavior of
quantum molecules \cite{pere}.

 The wave packets are sums over energy eigenstates $\psi_{r}(x)$
 ($x \in R^{n}$; here and later on we avoid the vector notation and
 omit vector indices if there is no danger of confusion).
  For anharmonic oscillators
 the leading behavior for large $|x|$ of $\psi_{r}(x)$
 is determined by the behavior of the ground state $\chi$ \cite{Simo}.
 We write the ground state in the form $\chi \equiv exp(- \frac{S}{\hbar})$ .
 It is the solution of the Hamiltonian
eigenvalue problem with the eigenvalue $E_{q}$
 \begin{equation} H \chi \equiv (-\frac{\hbar^{2}}{2m} \triangle + V)\chi
  = E_{q} \chi
  \end{equation}
 Let us make the similarity transformation
 \begin{equation}\check{H} \equiv \hbar^{-1} \chi ^{-1} H \chi =
  - \frac{\hbar}{2m}
 \triangle + \frac{1}{m} \nabla S \nabla \end{equation}
 Then, $\check{H}$ has a formal limit when $ \hbar \rightarrow 0 $
  equal to $\frac{1}{m}
 \nabla S \nabla $ (note that $S$ is real if we restrict ourselves
  to systems with a non-degenerate
 ground state $\chi$ ).

Let $\cal P$ be an analytic function on the configuration space.
Let us consider the Heisenberg equations
of motion on a state $exp(-\frac{S}{\hbar}) \cal P$ ( me mark position
and momentum operators by hats in order to distinguish them from an argument of
a wave function)
\begin{equation}-m \frac {d \hat{x}_{t}}{dt}
exp(-\frac{S}{\hbar}){\cal P} =\hat{p} _{t} exp(-\frac{S}{\hbar}){\cal P}
\end{equation}
where
\begin{equation} \hat{x}_{t}=U_{t}\hat{x}U_{t}^{+}\end{equation}
Using eq.(4) we can rewrite eq.(5) in the form
\begin{equation}-m\frac{d\hat{x}_{t}}{dt}exp(-\frac{S}{\hbar}){\cal P}
=exp(-\frac{S}{\hbar})\check{U}_{t}(-i\hbar \nabla
+ i \nabla S ) \check{U}_{t}^{+}{\cal P} \end{equation}
where $\check{U_{t}}$ is the unitary group generated by $\check{H}$
(eq.(4)).
The limit $\hbar \rightarrow 0$ in eq.(7) can be obtained explicitly.
This is so because $ \hbar \nabla$ acting on an
 analytic function of $ \hbar $ gives a function of order $ \hbar $.
Then,in the limit $\hbar \rightarrow 0$
\begin{equation}\check{U_{t}} {\cal P}(x)=
{\cal P}(x_{t}(x))+O(\hbar)\end{equation}
where $x_{t}(x)$ is the solution of the c-number equation
\begin{equation}m\frac{dx_{t}}{dt}=-i\nabla S(x_{t})\end{equation}
 with the initial condition x.

Neglecting terms of order $ \hbar $ we can see that equation (7)
 is simplified to
\begin{displaymath}-m\frac{d\hat{x_{t}}}{dt}exp(-\frac{S}{\hbar}){\cal P}=
exp(-\frac{S}{\hbar})(i\nabla S(x_{t})){\cal P}
\end{displaymath}
Hence, by iteration we obtain the equality
\begin{equation}f(\hat{x_{t}}){\cal P} exp(-\frac{S}{\hbar})=
 f(x_{t}) {\cal P} exp(-\frac{S}{\hbar}) \end{equation}
for an arbitrary analytic function f, where $x_{t}$ (without the hat)
 is the solution of the c-number differential equation (9).

The definition of the coherent state $|z>$ is now determined by the
requirements

i)$|z=0>=\chi$

ii) $|z>$ is transformed into a certain $|z(t)>$
 under the quantum Hamiltonian evolution (8) in the limit $\hbar
\rightarrow 0 $.

For the construction of $|z>$ let us consider a solution $S^{I} _{cl}(x)$
of the "imaginary" Hamilton-Jacobi equation (outside the classical
domain)
 \begin{equation} - \frac{1}{2m} (\nabla S^{I} _{cl})^{2} + V =
  E_{cl}(I)
   \end {equation}
 depending on n parameters $I_{k}$ which can be related to the action
 variables.We assume that $E_{cl}\equiv E_{cl}(I=0)$ is the classical
 ground state energy. It is known that $E_{q}-E_{cl}\simeq O(\hbar)$.
We denote by $S_{cl}$ the solution of eq.(11) with the ground state
  energy $E_{cl}$. The difference between $S$
  and $S_{cl}$ is negligible in the limit $\hbar \rightarrow 0$.
  This can be seen from equation (3) which when expressed by S reads
 \begin{equation} - \frac{1}{2m} (\nabla S)^{2} +
 \frac{\hbar}{2m}\triangle S + V = E_{q}
  \end {equation}
Subtracting eqs.(11) and (12) we can conclude that
 $S-S_{cl}\simeq O(\hbar)$.

  The search for a state $|z>$ fulfilling our assumptions i)-ii) is
  simplified if we find operators $Q_{k}$ which under the quantum evolution
$exp(-it\check{H})$ transform multiplicatively
 \begin{equation} Q_{k}|\psi> \rightarrow \lambda_{k}(t) Q_{k}|\psi>
  \end{equation}
 on a dense set of vectors $|\psi>$.
 We find that the proper guess is
 \begin{equation} Q_{k} = exp(\frac{\partial S^{I} _{cl}}
 {\partial I_{k} } )_{|I=0}
  \end{equation}
 From eq.(8) and the estimate on $S-S_{cl}$ it follows that $\check{H}$
 in the limit
  $\hbar \rightarrow 0$ generates the complex classical  evolution
 \begin{equation} \frac{ d\xi}{dt}=-\frac{i}{m} \nabla S_{cl}(\xi)
\end{equation}
  We show that in this limit the variables
 $ Q_{k}(t) $ fulfil linear equations. In fact, from eq.(14)
 \begin{equation} \frac{dQ_{k}}{dt} (\xi_{t}) = \frac{-i}{m} \frac{
\partial^{2}
  S_{cl}}{\partial I_{k} \partial \xi_{r}} \frac {\partial S_{cl}}
  {\partial \xi_{r}} Q_{k} (\xi_{t})\end{equation}
 Then, from eq.(11) it follows that
 \begin{equation}- \frac{1}{m} \frac{\partial^{2} S_{cl}}
 {\partial x_{r}\partial
 I_{s}} \frac{\partial S_{cl}}{\partial x_{r}} =
  \frac{\partial E_{cl}}{\partial I_{s}}
  \end{equation}
  Hence, setting $I=0$ in eq.(17)
 \begin{equation}\frac{dQ_{k}}{dt}(\xi) =-i\omega_{k}Q_{k}(\xi)\end{equation}
 where
 \begin{equation}
 \omega_{k}= \frac{\partial E_{cl}}{\partial I_{k}}(I=0)
 \end{equation}
 It follows that in eq.(13) $\lambda_{k}(t)=exp(-i\omega_{k} t)$.

    Now, in terms of the variables $Q_{k}$ our definition of
     a coherent wave packet reads

 \begin{equation} <x\mid z>\equiv exp(-\frac{S}{\hbar}+ \sqrt{\frac{2}
 {\hbar}}\sum_{k=1}^{n}\sqrt{\omega_{k}}z_{k}Q_{k} )\end{equation}
 The linear dependence on Q of the exponential in eq.(20) has as a consequence
 that $|z>$ is an eigenfunction of a generalized annihilation operator
 \begin{equation}
 \alpha_{k}=\sqrt{\frac{\hbar}{2\omega_{k}}}\frac{\partial}{\partial Q_{k}}+
 \frac{1}
 {\sqrt{2\hbar\omega_{k}}}\frac{\partial S}{\partial Q_{k}}
 \end{equation}
 Namely
 \begin{displaymath}
 \alpha_{k}|z>=z_{k}|z>
 \end{displaymath}
 It follows that
 \begin{equation}< z\mid \frac{1}{2}(\alpha_{k}+\alpha_{k}^{+})\mid z>
 <z\mid z>^{-1}=Rez_{k}\end{equation}
 and
 \begin{equation}< z\mid \frac{1}{2i}(\alpha_{k}-\alpha_{k}^{+})
 \mid z>< z\mid z>^{-1}=Imz_{k}\end{equation}

 In this way we obtain an interpretation of $z$ as a mean value of a certain
 observable.Furthermore, the eigenvalue condition is a necessary condition for
 a minimal uncertainty property of the operators which constitute the real
 and imaginary part of $\alpha$ (see refs.\cite{jack} and \cite{carr}).
 It follows from eqs.(8),(18) and (20) that under the quantum evolution
 \begin{displaymath}
 U_{t}|z>=|exp(-i\omega t)z> + O(\hbar)
  \end{displaymath}

  There remains to investigate the
  localization properties of $|z>$.The probability density
  $|<x|z>|^{2}$
 is maximal at a point Q determined by the equation
 \begin{equation} \frac{\partial S}{\partial Q_{r}} = \sqrt{2\hbar\omega_{r}}
  Rez_{r}   \end{equation}
 Eq.(24) gives an interpretation of Rez which coincides with the one of
  eq.(22) in the leading order of $\hbar$.The probability density
  $|<x|U_{t}|z>|^{2}$
 is concentrated on the solution of the equation
 \begin{equation} \frac{\partial S}{\partial Q_{r}} = \sqrt{2\hbar\omega_{r}}
  Re(exp(-i\omega_{r}t)z_{r})   \end{equation}
  It can be seen from the definition (14) of Q  that $Q\sim x$ and $S(Q)\sim
Q^{2}$ for the
  anharmonic oscillators in the weak coupling regime.
  So, in this sense the probability density $|<x|U_{t}|z>|^{2}$ is concentrated
on a
  deformed torus filled by the trajectories of the integrable system.

Let us consider two examples.The first concerns a onedimensional
non-negative potential V normalized in such a way that $E_{cl}=0$ in eq.(11).
 Then
\begin{equation} S_{cl}(x)=\sqrt{2m}\int^{x}\sqrt{V(y)}dy
\end{equation}
is uniquely defined (no problem with the square root ).Hence
\begin{equation}Q = exp(\omega \sqrt{m} \int^{x}\frac {1}
{\sqrt{2V(y)}}dy )\end{equation}
To be more specific let
\begin{equation}V(y)=\frac{m\omega^{2}}{2}y^{2} + g y^{4}
 \end{equation}
We calculate the integral (26)

  \begin{equation} S_{cl}(x)=m\omega \int^{x}dy y\sqrt{1 + cy^{2}}=
 \frac{m\omega}{3c}(1+cx^{2})^{\frac{3}{2}}\end{equation}
 where $ c=\frac{2g}{m\omega^{2}} $.
 Now,
 \begin{equation}Q=exp(\int^{x}dy(y\sqrt{1+cy^{2}} )^{-1})=
 2x(1+\sqrt{1+cx^{2}} )^{-1}\end{equation}

As a second example let us consider a non-negative twodimensional rotationally
symmetric
potential V(r).Then, the classical Hamilton-Jacobi function W reads (in the
plane
coordinates $(r,\phi)$)
\begin{equation}
W(r,\phi)=I_{1}\phi+\int^{r} d\rho(2mE_{cl}(I_{1},I_{2})-2mV(\rho)-I_{1}^{2}
\rho^{-2})^{\frac{1}{2}}
\end{equation}
A solution of the "imaginary" Hamilton-Jacobi equation
(11) is obtained in the limit $E_{cl}(I_{1},I_{2}) \rightarrow 0$.
For a non-negative potential
this limit means $I_{1}\rightarrow 0$ and $I_{2}\rightarrow 0$. It follows that
\begin{equation}
S_{cl}(r,\phi)=\int^{r} d\rho(2mV(\rho))^
{\frac{1}{2}}
\end{equation}
and
\begin{equation}
Q_{1}=exp(i\phi+\omega_{1}u(r))
\end{equation}
\begin{displaymath}
Q_{2}=exp(2\omega_{2}u(r))
\end{displaymath}
where
\begin{equation}
u(r)= \int^{r} d\rho(2mV(\rho))^{-\frac{1}{2}}
\end{equation}
and $\omega_{k}$ are defined in eq.(19).

So far we have restricted ourselves to the limit $\hbar \rightarrow 0$ .
The corrections to the formulae (8)-(10) and (13) are discussed in our papers
\cite{hab1}\cite{hab2}.In particular, in ref.\cite{hab2} we bound the growth
in time of the $O(\hbar)$ term in eq.(8) by $\sqrt{\hbar t}$. We specify
also some conditions  which ensure that
the $O(\hbar)$ term is bounded uniformly in time. On the basis of these results
we estimate that the wave packet moves along the trajectory (25) and
does not change its shape at least for a time of order $\frac{1}{\hbar}$.
In general, it is known \cite{pere} (see also
a rigorous formulation in ref.\cite{ruel}) that a particle in a state
belonging to the subspace of the Hilbert space corresponding to the discrete
spectrum remains localized in a bounded region for an arbitrarily large time.
Moreover, from the quasiperiodicity it follows that the wave packet
recovers its shape after a sufficiently large time (see an estimate on this
time in ref.\cite {pere}).Our claim is that the coherent states defined in
this Letter do not lose their shapes during a time interval large in comparison
to the classical periods $2\pi\omega_{k}^{-1}$.

\end{document}